\documentclass[11pt]{article}
\usepackage[a4paper,bindingoffset=0.in,left=2.5cm,right=2.5cm,top=2cm,bottom=3cm]{geometry}
\usepackage{amsmath, amssymb,amsthm}
\usepackage{graphicx,subfigure}
\usepackage{color}
\usepackage{url}
\usepackage[a]{esvect}
\usepackage{enumitem}
\usepackage{lmodern}
\usepackage{setspace}

\usepackage[colorinlistoftodos]{todonotes}

\begin{document}
\title{Understanding Complex Systems: \linebreak From Networks to Optimal Higher-Order Models}

\author{Renaud Lambiotte$^{\dagger}$, Martin Rosvall$^{\star}$, Ingo Scholtes$^{\ddagger}$ \\
{\small $^{\dagger}$ University of Oxford, United Kingdom} \\
{\small $^{\star}$ Integrated Science Lab, Ume\aa{} University, Sweden }\\
{\small $^{\ddagger}$ Chair of Systems Design, ETH Z\"urich, Switzerland} 
}
\date{}
\maketitle

\begin{abstract}
To better understand the structure and function of complex systems, researchers often represent direct interactions between components in complex systems with networks, assuming that indirect influence between distant components can be modelled by paths. Such network models assume that actual paths are memoryless. That is, the way a path continues as it passes through a node does not depend on where it came from. Recent studies of data on actual paths in complex systems question this assumption and instead indicate that memory in paths does have considerable impact on central methods in network science. A growing research community working with so-called higher-order network models addresses this issue, seeking to take advantage of information that conventional network representations disregard. Here we summarise the progress in this area and outline remaining challenges calling for more research.
\end{abstract}

A long-standing goal of statistical physics is to understand emergent phenomena in complex systems that consist of a large number of interacting components.
Such systems not only occur in condensed matter physics, but they are also widespread in other disciplines, and physicists have been able to contribute to a better understanding of biological, social, economic, and technological systems. 
A salient feature of complex systems is that all system components can influence each other, either directly or indirectly.
Systems for which the topology of these interactions is unknown are often studied using mean-field approaches, which summarise interactions between all elements with a single averaged field.
Over the past few decades, a surge of data has demonstrated that complex systems in the real world exhibit sparse and complex topologies in which few components directly interact with each other, while most components indirectly influence each other via sequences of multiple direct interactions~\cite{Newman2003}.
Such systems can be conveniently represented as \emph{graphs} or \emph{networks}, where nodes $x_i$ represent the components of a system and \emph{links} $\vv{x_ix_j}$ capture the topology of direct pairwise interactions.
The indirect influence between two components $x_0$ and $x_n$ can be studied based on sequences of direct interactions $\vv{x_0x_1}, \vv{x_1x_2}, \ldots, \vv{x_{n-1}x_n}$ that mediate the influence between $x_0$ and $x_n$ via a \emph{path} $\vv{x_0\ldots x_n}$.

Building on this abstraction, \emph{network science} has developed methods that help us to better understand the structure and function of complex systems.
The success and popularity of these network science methods across disciplines rest on their broad applicability to relational data that capture pairwise interactions.
However, the analysis of such data based on linear transformations, algebraic methods, and Markovian models of dynamical processes~\cite{Boccaletti2006,Porter2016} also makes an important assumption, namely that the \emph{paths} by which a system's components \emph{indirectly} influence each other can be understood based on the \emph{transitive closure} of direct, pairwise interactions.
That is, most network science methods rest on the assumption that the existence of direct interactions $\vv{x_0x_1}$ and $\vv{x_1x_2}$ implies that $x_0$ can indirectly influence $x_2$ via a transitive path $\vv{x_0x_1x_2}$.
Notably, methods based on the composition of linear transformations that capture network topologies implicitly introduce this fundamental hypothesis about indirect influence in a complex system.
Examples include algebraic and spectral methods based on eigenvalues, products, and powers of adjacency matrices and Laplacians; Markovian models of dynamical processes; or algorithmic techniques that calculate flows and paths in graphs.

The transitive path hypothesis can be justified if our knowledge about a system is limited to pairwise interactions.
Thus, we should view a network model as a \emph{maximum entropy model of paths}~\cite{jaynes1957information,franccoisse2017bag} in a complex system, much like a mean-field model can be seen as a maximum entropy model for the topology of direct interactions. 
From a statistical point of view, this interpretation of a network model corresponds to the assumption that the links $\vv{x_ix_{i+1}}$ that contribute to a path $\vv{x_0x_1\ldots x_n}$ are statistically independent or, equivalently, that the sequences of nodes traversed by paths are memoryless, first-order Markov sequences.
The question of whether this strong assumption can be justified in real systems is at the heart of a growing research community centred on higher-order network models. 
Building on advances in data collection and sensing techniques, this community uses new forms of high-dimensional sequence or time-series data to reconstruct the complex paths of interactions in complex systems. Examples include time-series data capturing time-stamped interactions in social networks~\cite{Karsai2012,Pfitzner2013,Wei2015}, temporal patterns in trade relations~\cite{Lentz2013,Koher2016,Lentz2013}, navigation paths of humans in information networks~\cite{asztalos2010network,backstrom2011supervised,singer2014memory,Scholtes2017}, and traces of dynamical processes in networked systems~\cite{Xu2016}. Similar to how the surge of data on direct interactions has questioned mean-field models, these studies of actual paths in complex systems show that standard network models can blur our understanding of critical nodes and functional modules~\cite{Scholtes2014,Rosvall2014,Peixoto2017}.

After briefly introducing the higher-order network model framework, this progress article provides examples where higher-order network models offer new and critical insights into the systems that they represent by asking:
How do path structures in complex systems affect the applicability of popular network science methods?
How do higher-order network models help us to improve our understanding of complex systems? 
And which complex systems require higher-order modelling techniques?
Specifically, we address these questions in light of three foundational areas of network science: community detection, ranking nodes, and modelling dynamical processes.
The answers present both challenges and opportunities for developing new data-driven modelling techniques to better understand complex systems.

\paragraph{Higher-order models of interactions in complex systems}
 
Owing to the different systems and data under study, the growing community working with higher-order network models has taken a multi-pronged approach:
One stream of works builds on the observation that many systems exhibit \emph{many-body interactions}, which require generalising \emph{links} to higher-order models that capture more than just pairwise interactions~\cite{arenas2008motif,Petri2013,Benson2016}.
Examples include triangles that are known to be fundamental building blocks of social networks~\cite{granovetter1977strength}, cliques in scientific coauthorship networks~\cite{patania2017shape}, feed-forward loop network motifs in biochemical transcription networks~\cite{mangan2003structure}, and spatial coexistence relations between species in an ecosystem~\cite{Levine2017}, as well as \emph{trigenic interactions} in gene regulatory networks~\cite{kuzmin2018systematic}.
Another line of research acknowledges that the links in many complex systems can be of multiple types, which calls for multi-layer generalisations of network models~\cite{kivela2014multilayer,de2016physics}.
Examples include multi-modal transportation systems~\cite{cardillo2013modeling}, interdependent layers of power and communication infrastructures~\cite{buldyrev2010catastrophic}, and multi-layer financial networks~\cite{battiston2016financial}.
And finally, the increasing quantity of high-resolution time-series data can be used to infer temporal sequences or paths of interactions between more than two components in a system.
Examples include data on scholarly citation networks~\cite{Rosvall2014}, time-stamped social networks~\cite{Holme2012,Holme2012}, and patient pathways in hospital networks~\cite{palla2017complex}, as well as human trajectories in transportation and information networks~\cite{Scholtes2017}.

\begin{figure}[htb]
\centering
\includegraphics[width=0.9\columnwidth]{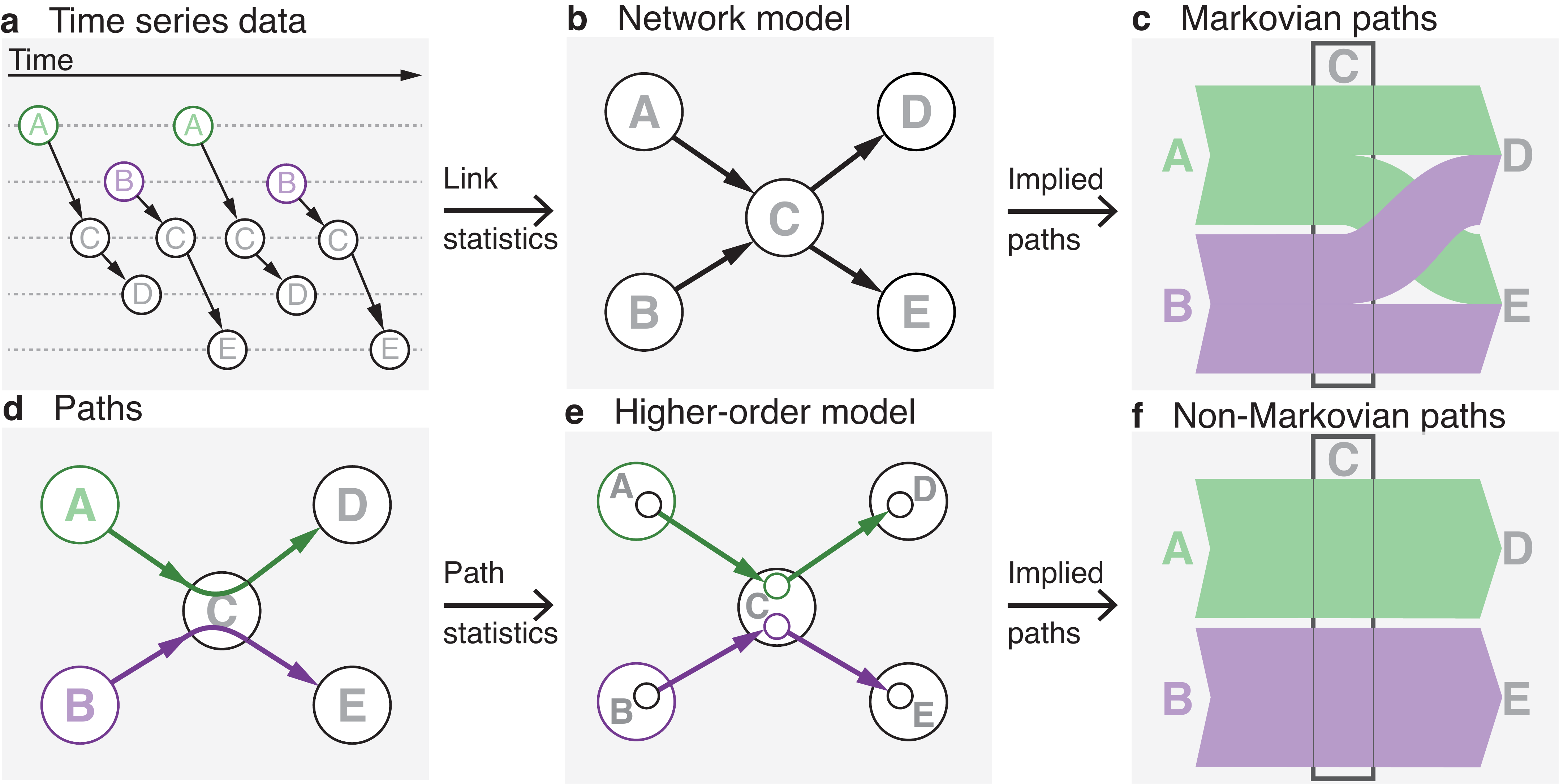}
\caption{\label{memorynetwork} Higher-order models better capture the topology of paths in complex systems. A rich source of path information is time-series data that capture interaction sequences between a system's components (a). Focusing on pairwise interactions, network models abstract a system's topology with nodes and links (b) while assuming that paths are transitive and Markovian (c). Due to the chronological ordering of interactions, the actual paths of indirect influence between system components (d) can deviate from this assumption. Focusing on \emph{multi-step paths} rather than \emph{pairwise interactions}, higher-order network models (e) can capture the actual topology of indirect influence, allowing to apply network methods to systems with non-Markovian paths (f).}
\end{figure}

Despite differences in motivation and mathematical underpinning, these approaches have one aspect in common:
They account for the fact that standard network models are too simple to explain the complex paths of influence captured in the growing volume of rich data on biological, technical, economic, and social systems.
Figure \ref{memorynetwork} illustrates this in time-series data.
Such data often provide information on real-world paths, either directly, by capturing sequences of interactions by which processes or cascades affect nodes, or indirectly, by capturing time-stamped links that can be used to reason about \emph{causal} or \emph{time-respecting paths}~\cite{Holme2012}.
Focusing on pairwise interactions, a standard network model represents the link topology of the underlying system (Fig.~\ref{memorynetwork}b).
This representation discards information on how links contribute to paths, implicitly implying that nodes can indirectly influence each other via transitive paths that traverse nodes in a memoryless, Markovian fashion.
In our example, nodes $\mathsf{A}$ and $\mathsf{B}$ can both indirectly influence $\mathsf{D}$ and $\mathsf{E}$ via four transitive paths $\vv{\mathsf{ACD}}$, $\vv{\mathsf{ACE}}$, $\vv{\mathsf{BCD}}$, and $\vv{\mathsf{BCE}}$ (Fig.~\ref{memorynetwork}c).
A closer look at the ordering of interactions in the time-series data reveals that only two of these four possible paths exist in the sequence (Fig.~\ref{memorynetwork}d), thus invalidating network analytic methods that assume transitive, Markovian paths.
Overcoming this shortcoming requires a \emph{path-centric} view of data that generalises networks to higher-order models of paths~\cite{Rosvall2014,Scholtes2014,Scholtes2016,Xu2016,Scholtes2017,edler2017mapping}.
Figure~\ref{memorynetwork}d illustrates this idea with a second-order model that accounts for the topology of paths of length two.
In the spirit of higher-order Markov chain models, this model can be represented with a \emph{memory network}~\cite{Rosvall2014}, where \emph{state nodes} represent states in a second-order state space and links encode possible transitions between states.
Depending on the topology of paths, each of the five \emph{physical nodes} $\mathsf{A}$, $\mathsf{B}$, $\mathsf{C}$, $\mathsf{D}$, and $\mathsf{E}$, which typically are the object of interest in the real world, has one or more state nodes.
These state nodes enable efficient higher-order network models of paths: A path described by a Markovian model on the state nodes, which directs the path from one state node to the next with a probability that does not depend on previously visited state nodes, appears \emph{non-Markovian} on the physical nodes (Fig.~\ref{memorynetwork}f).
This modelling approach can be generalised to arbitrary orders $m$ by adding one state node for each prefix of $m-1$ nodes that precede the current physical node on a path.
In this way, we can construct network models that capture higher-order effects in paths for any given order $m$.

\begin{figure}[htbp]
\centering
\includegraphics[width=0.8\columnwidth]{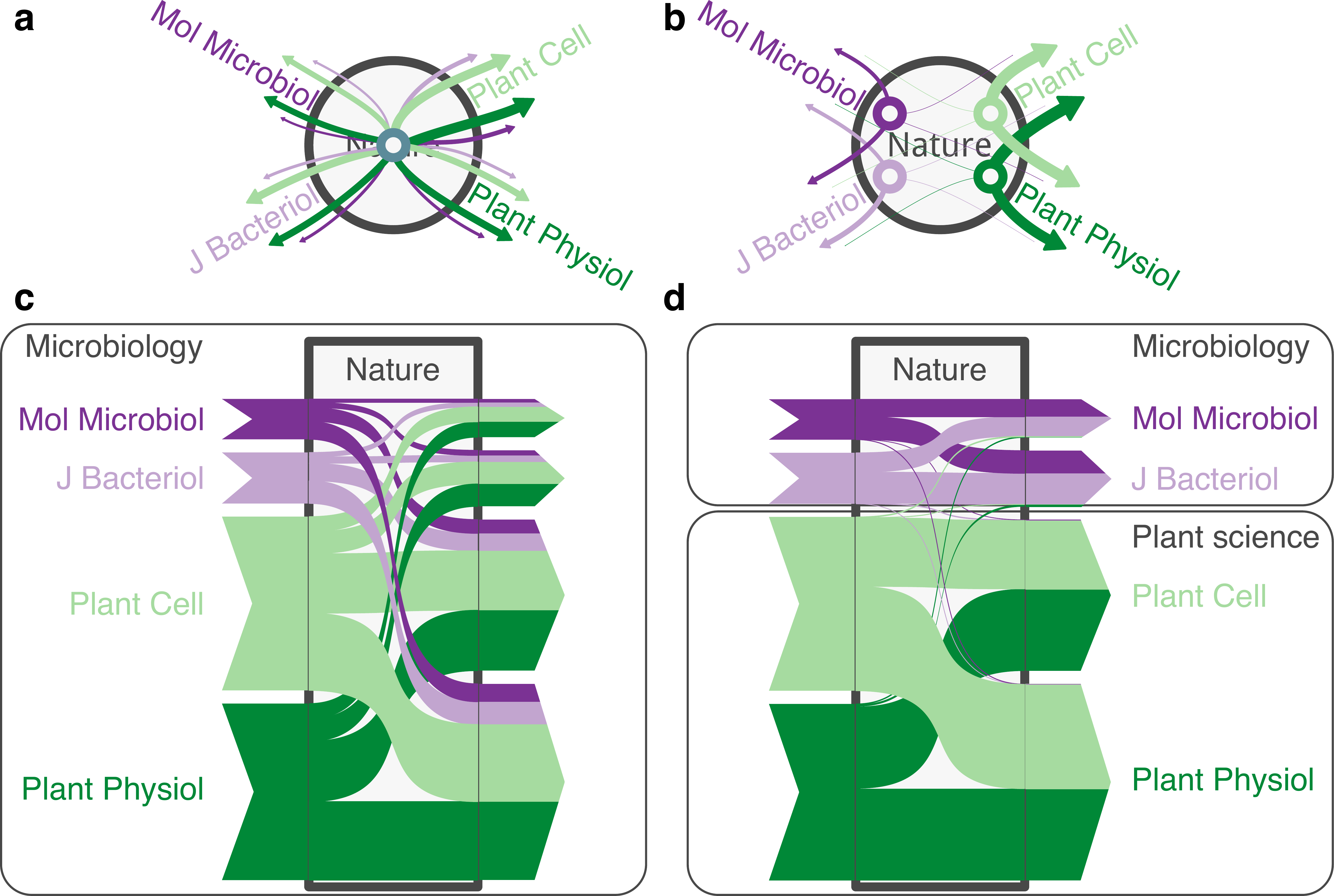}
\caption{\label{citationflows} Community detection of paths can capture overlapping communities. (a) A conventional first-order Markov representation of citation flows from four specialised journals through multidisciplinary Nature. (b) A second-order representation with one state node for each citing journal. (c) The conventional network representation mixes flows and washes out the boundary between fields. (d) A second-order Markov model captures the fact that citation flows through a multidisciplinary journal depends on where they come from and allows overlapping fields in Nature. }
\end{figure}

\paragraph{Non-Markovian paths and community detection}
Community detection~\cite{Fortunato2010} is an umbrella term for a large number of algorithms that group nodes into modules to simplify and highlight essential structures in the network topology. 
Since higher-order representations can capture more complex forms of interactions, community detection algorithms generalised to higher-order representations can capture more complex forms of relational regularities. For example, for citations flows between journals and communities with long flow persistence times, a standard network model with weighted directed links between journals aggregated from citations between their articles fails to capture the complex citation flows through multidisciplinary journals such as Nature (Fig.~\ref{citationflows}a,c)~\cite{Rosvall2014}. Independent of where the flows come from, they continue to another journal proportionally to the number of citations to that journal among all articles published in the multidisciplinary journal. As a consequence, citation flows from different fields mix and move in a non-realistic way from one field to another. For example, Fig.~\ref{citationflows}c illustrates how most citation flows from two microbiology journals continue to two plant science journals such that all these journals are best assigned to the same field. Accordingly, community detection based on a standard network model can wash out boundaries between modules and fail to assign nodes to multiple overlapping modules.

In contrast, a second-order Markov representation of citations flows, which takes into account where citations come from, captures the fact that most citation flows coming to Nature from one field return to the same field (Fig.~\ref{citationflows}b,d). For example, when going from a first- to a second-order Markov representation, the relative amount of citation flows that return back to the same journal after two steps, averaged over all journals, increases from 11\% to 22\%~\cite{Rosvall2014}. Moreover, the citation flows that do not return move in more realistic ways: Figures \ref{citationflows}b and d illustrate how citation flows from the Journal of Microbiology and the Journal of Bacteriology in microbiology mostly return to either journal and, similarly, how citation flows from Plant Cell and Plant Physiology in plant science mostly return to those journals. As a consequence, citation flows stay within their respective fields, highlighting the multidisciplinary character of the journal Nature. 
Assigning the state nodes of a multidisciplinary journal to different fields better captures the citation flows in the second-order Markov representation. In this way, a second-order representation of non-Markovian citation paths is critical for capturing overlapping research fields in multidisciplinary journals. 

\paragraph{Non-Markovian paths and node centralities}
The development of algorithms to identify important nodes is one of the major success stories in network science.
They help us to locate critical elements in networked infrastructures, identify influential actors in social systems, or find relevant pages in the World Wide Web.
At the heart of these applications are measures for the \emph{centrality of nodes}, for example, using their occurrence on the shortest paths between other nodes, their role in flow processes, or their influence on the steady state of dynamical processes~\cite{Newman2003,borgatti2005centrality,brandes2017positional}.
These methods assume that node centrality can be characterised based on the topology of pairwise interactions between system components.
However, in social systems, information is shared via specific paths that are influenced by context, preferences, and time and that cannot be understood merely based on the network topology.
Similarly, the topology of hyperlinks is not enough to explain the complex paths by which humans navigate the World Wide Web.
And in biological systems, the topology of pairwise interactions in a cell is hardly sufficient to understand the complex paths by which proteins, neurons, or genes influence each other.

\begin{figure}[t]
\centering
\includegraphics[width=.7\columnwidth]{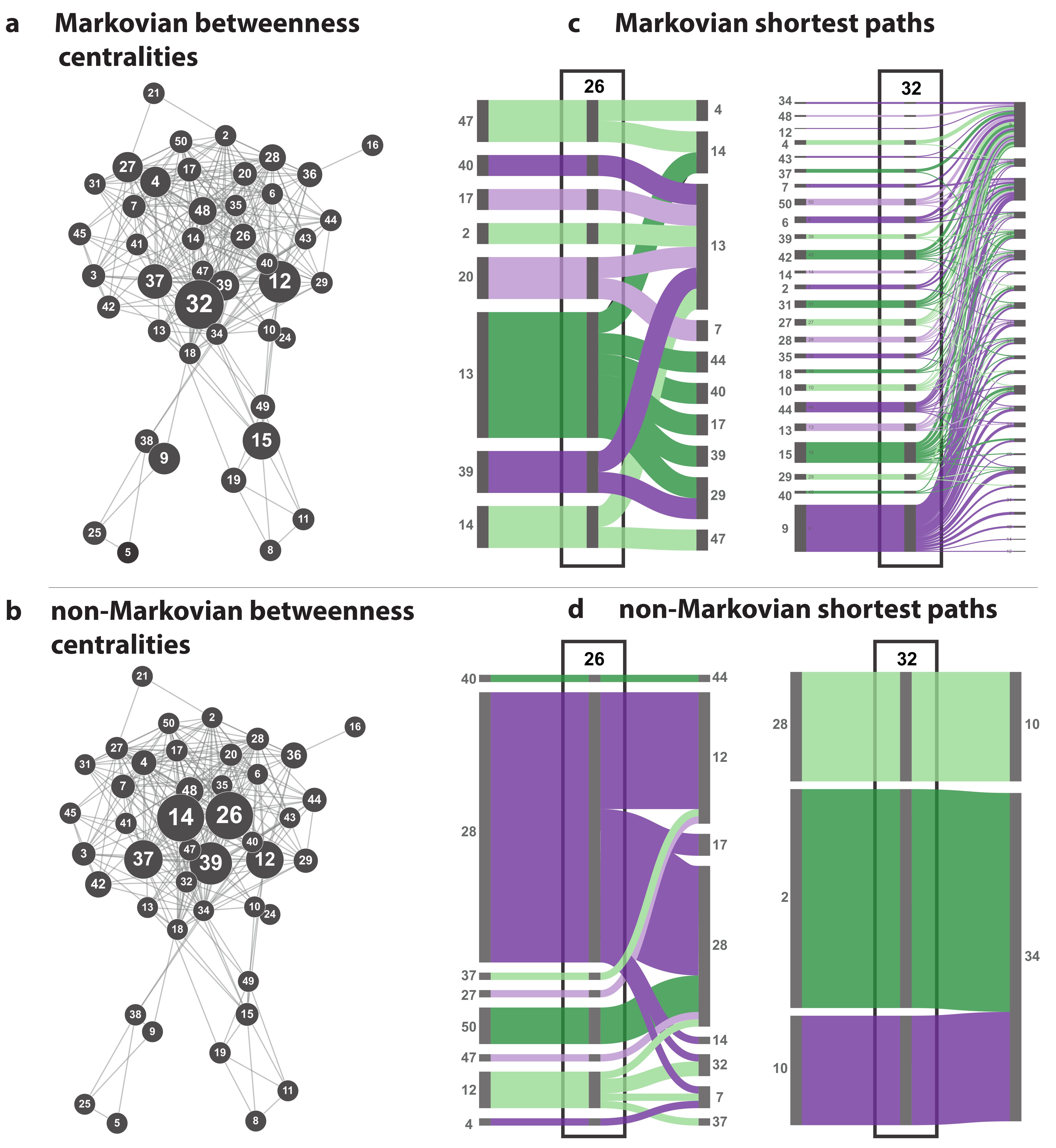}
\caption{\label{socialnetwork} Node centralities calculated based on shortest paths in a network model of time-stamped social interactions (a) do not capture the true importance of nodes calculated based on causal paths that respect causality in the underlying time-series data (b). The alluvial diagrams highlight the fact that the chronological order of interactions alters the shortest \emph{causal paths} passing through nodes 26 and 32 (d), compared to what we expect based on the topology of direct interactions (c), thus considerably changing the betweenness centrality of nodes.}
\end{figure}

To truly capture the importance of nodes, we must go beyond network models and account for the complex structure of paths in high-dimensional, time-resolved data.
An example is shown in Figure~\ref{socialnetwork}, which is based on time-stamped social interactions between the developers in a major Open Source Software community.
The network model of these interactions (Fig.~\ref{socialnetwork}a) allows us to estimate the importance of nodes, for example, using \emph{betweenness centrality}, a measure that assigns high centrality to a node $v$ if many shortest paths between pairs of other nodes pass through $v$.
The resulting node centralities are represented by node sizes in Fig.~\ref{socialnetwork}a (bottom), indicating that node 32 is the most important node in the system.

But is this a good estimate for the actual importance of developers?
We can answer this question by inferring \emph{causal paths} in the underlying time-series data That is, we consider which paths actually exist based on the chronological ordering and timing of time-stamped interactions.
In a nutshell, for a sequence of two interactions $\vec{ab}$ and $\vec{bc}$, a \emph{causal path} $\vec{abc}$ only exists if $\vec{ab}$ occurs before $\vec{bc}$. 
Hence, time-stamped network data allow us to calculate causal path statistics that may or may not be consistent with the assumption of transitive, Markovian paths that are implicitly made when using a network model~\cite{Scholtes2016}.
In the example shown in Fig.~\ref{socialnetwork}, a calculation of betweenness centralities based on actual shortest \emph{causal paths} considerably shifts the importance of nodes. 
The alluvial diagrams in Fig.~\ref{socialnetwork}a and b visualise these differences, revealing that the shortest causal paths passing through node 32 are considerably more constrained than expected.
This is due to complex temporal patterns in human communication behaviour that are not captured by a network model~\cite{Pfitzner2013}.
As a result, node 32 is less central than we would assume, based on the network topology.
In contrast, node 26, which ranks among the least central nodes from a topology perspective, turns out to be the most important node in terms of causal paths in the temporal interaction sequence.

Higher-order models open new ways to address these limitations of existing centrality measures.
We can, for instance, generalise path-based centrality measures like betweenness or closeness  to higher-dimensional \emph{De Bruijn} graphs, which like memory networks have nodes that represent sequences of length $m$ and links between sequences that overlap by $m-1$ components as between $\vv{x_0\ldots x_{m-1}}$ and $\vv{x_1\ldots x_{m}}$~\cite{deBruijn1946,Scholtes2016}.
Similarly, spectral measures such as PageRank or eigenvector centrality can be redefined based on eigenvectors of linear operators derived from De Bruijn graphs or memory networks~\cite{Rosvall2014,Scholtes2017,Xu2016}.
Such novel measures help us to quantitatively assess the true importance of elements in a complex system, considering a system's network topology as well as temporal patterns in non-Markovian paths.

\paragraph{Non-Markovian paths and dynamical processes}
Along with enabling us to reason about topological features such as community structures or node centralities, network science has improved our understanding of how a system's topology influences dynamical processes, and thus a system's \emph{function}.
Much of this research is based on the analytical study of linear (or linearisable) dynamical systems in which Laplacian, adjacency, or transition matrices encode the topology of direct interactions between the system's elements.
The fact that this allows us to treat processes in networks in much the same way as in continuous space -- with linear operators like the Laplacian matrix generalising differential operators to arbitrary discrete interaction topologies -- naturally appeals to physicists.
The eigenvalues and eigenvectors of these matrix operators capture the way the topology of a system influences the efficiency of the diffusion and propagation processes, whether it enforces or mitigates the stability of dynamical systems, and if it hinders or fosters collective dynamics.

\begin{figure}[tb]
\centering
\includegraphics[width=.9\columnwidth]{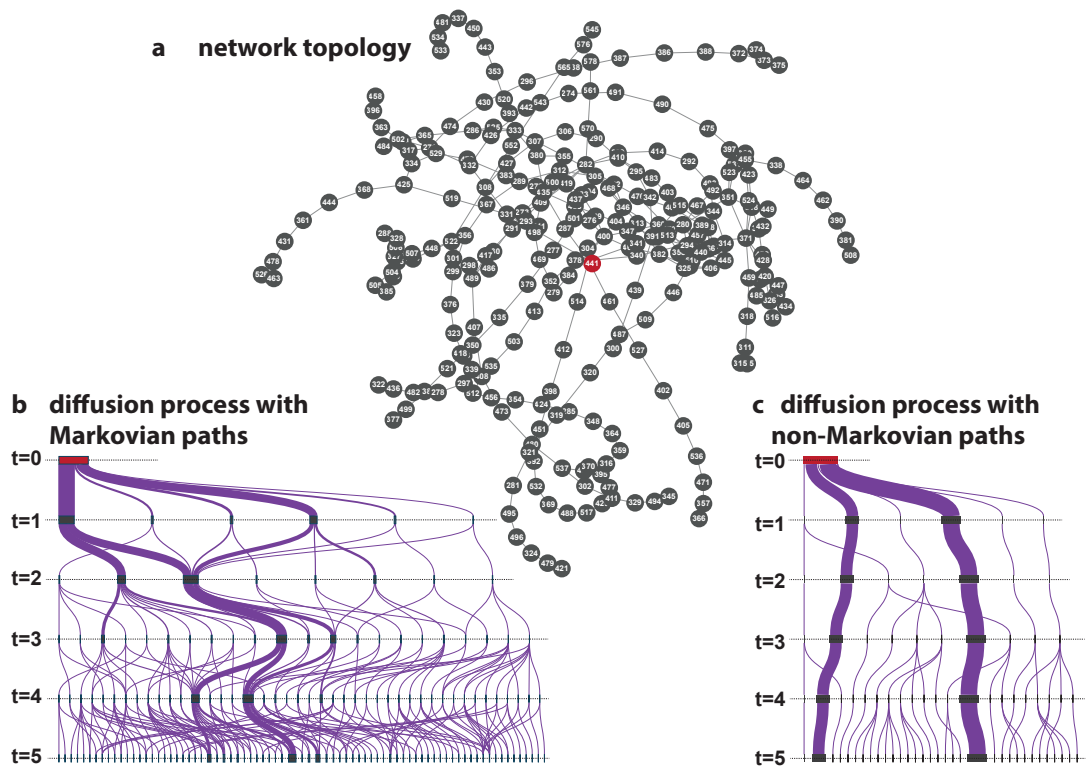}
\caption{\label{dynamics}Evolution of a diffusion process in a network model of the London Tube (a). The flow diagrams in (b,c) show the first five steps of a discrete-time diffusion process initiated in node 441 (marked in red). While (b) shows the dynamics of the process in the network topology, (c) shows the evolution of diffusion across the non-Markovian paths created by the specific ordering of train connections in the London Tube. The causal topology created by paths in real systems crucially influences dynamical processes and challenges our understanding of complex systems.}
\end{figure}

While it has provided powerful tools to relate the structure and dynamics of complex systems, the assumption of transitive, Markovian paths implicitly introduced by the application of algebraic methods is not justified in many real systems. 
Figure \ref{dynamics}a illustrates an example of such a complex system, which shows a network model of the London Tube, where the nodes are train stations and links capture direct train connections between stations.
To understand how the topology of this transportation network influences its efficiency and robustness, it is common to study its influence on dynamical processes.
As a simple example, let us consider a discrete-time model that captures the diffusion of passengers who start their journey at a single station at time $t=0$, travelling one station per discrete time step.
We further adjust each passenger's probability of continuing across a given link based on data on average passenger volumes between London Tube stations, making it more likely to continue through links with high passenger volume.
The flow diagram in Figure \ref{dynamics}b shows the first five steps in this process. Assuming transitive, Markovian paths, the diagram highlights how paths in the network topology shape the diffusion process.
However, real passenger paths are clearly not Markovian, thus bringing into question what the network topology can actually teach us about the properties of the real system.
Alternatively, using available data on actual passenger itineraries, we can study this diffusion process based on real paths (Fig.~\ref{dynamics}c).
This study reveals that neither the topology of the system nor the aggregate volumes of passengers travelling across links is sufficient to explain the complex non-Markovian paths and flows in the system.
Specifically, Fig.~\ref{dynamics}c reveals a strong directional preference that is rooted in the non-Markovian characteristics of paths and the underlying geography.
These patterns considerably change the evolution of dynamical processes in the system and limit what a system's topology can tell us about the robustness and efficiency of real transportation networks.

Higher-order models help us to overcome these limitations of standard network methods.
We can, for instance, generalise Laplacian and transition matrices to higher-dimensional \emph{De Bruijn} graph models~\cite{deBruijn1946,Scholtes2014} that capture a system's \emph{causal topology}, which is due to the non-Markovian characteristics of paths.
Such higher-order network representations are the basis by which we can apply methods from the study of dynamical systems, such as eigendecompositions, spectral analysis or stability theory, to systems with non-Markovian paths.
They allow us to analyse the complex interplay between time and topology in networked systems, and explain why non-Markovian characteristics of paths can both decelerate and accelerate dynamical processes and collective dynamics~\cite{Scholtes2014}.

\paragraph{Perspectives}

To explain the properties of complex systems, it is important to understand how a system's components influence each other.
Network science provides powerful computational and analytical tools to address this challenge based on network abstractions of direct, pairwise interactions.
This remarkably simple abstraction has helped us to uncover emergent phenomena that are tied  the essential features of the interaction topology rather than to the details of a given system.
Moreover, the combination of graph-theoretic methods with ensemble-based techniques from statistical physics has developed into an important foundation for statistical analysis, inference, and machine learning tasks in relational data.
However, thanks to rich data on social, technical, and biological systems, the limits of what network models can possibly teach us about real systems are becoming increasingly evident.
These data highlight the need for advanced modelling techniques that capture the complex \emph{non-Markovian interaction paths} observed in real systems.

An important epistemological challenge is to find new ways to infer optimal models of complex systems, given high-dimensional data.
Referring to Ockham's razor, such models should be maximally parsimonious. That is, we want to limit our assumptions to enable generalisable statements that go beyond the concrete system under study.
But a good model must still be complex enough to explain paths observed in real systems, which is where standard network models often fall short.
In other words, we are interested in higher-order models that \emph{compress} available information by modelling common patterns in paths, much in the same way as network science has exposed common topological patterns across a wide array of systems.
Finding optimal models based on rich data on complex systems thus turns into a machine learning problem, where standard network models are merely one of many possible explanations for observed paths.
Suitably adapted model selection and statistical learning techniques can be used to learn the optimal order of higher-order models from time-series data~\cite{Scholtes2017}, but little is known about how this challenge can be solved for other types of models and data.

The fact that the size of higher-order models can grow quickly as model dimensionality increases introduces both statistical and scalability challenges that must be addressed.
Reliable inference of models with high orders potentially requires vast volumes of data that may not be available in many settings.
Moreover, the use of fixed, single-order models can introduce the risk of simultaneously under- and overfitting patterns in path data and unnecessarily complicate higher-order models.
This highlights the need for computational and statistical methods that utilise variable-order~\cite{persson2016maps,Xu2016} and multi-order models~\cite{Scholtes2017}, as well as model order reduction techniques~\cite{Peixoto2017}, to generate computationally tractable models that neither under- nor overfit the data.

In addition to these methodological issues, the study of non-Markovian paths also foreshadows a new class of \emph{higher-order generative network models} similar to, for example, the stochastic block model~\cite{snijders1997}, the Watts-Strogatz model~\cite{watts1998} or the Barab\'asi-Albert~model~\cite{barabasi1999}.
In network science, such generative models play a crucial role both in the detection of structural patterns in networks and in the identification of maximally simple mechanisms from which they emerge.
Little is known about the mechanisms by which similar non-Markovian patterns emerge across different systems, a potentially fruitful research area.

Finally, there are broad perspectives for research on the relationship between higher-order network models and dynamical processes.
While most research has focused on linear diffusive processes, some works have considered the impact of non-Markovian paths on metapopulation models, akin to reaction-diffusion on networks \cite{belik2011natural,poletto2013human}. 
The properties of general non-linear processes with non-Markovian interaction paths are mostly unexplored.
These include coupled dynamical systems, for example, oscillators, on higher-order networks,  the identification of conditions that allow local and non-local patterns to emerge when standard tools such as master stability analysis are not applicable \cite{barahona2002synchronization}, and the use of higher-order models to study network control~\cite{Zhang2017b}.

Higher-order modelling techniques allow us to leverage existing network methods, extending them toward optimal models in order to explain the interaction topology in complex systems.
Such models and techniques not only create new opportunities for interdisciplinary exchanges between physics, computer science, and statistics, they can also help to reconcile the complex systems community with those researchers who have voiced concerns about the -- at times naive -- application of maximally simple, physics-inspired models.

\bibliographystyle{naturemag}
\bibliography{bibliography}

\end{document}